\documentstyle[itw991]{article}
\isitAfour
\begin{document}
\isittitle{ Improvement of SNR with Chaotic Spreading Sequences for CDMA}
\isitauthor{Ken Umeno\footnotemark  Ken-ichi Kitayama}
{Communications Research Laboratory, 
Ministry of Posts and Telecommunications\
4-2-1,Nukui-Kitamachi,Koganei,
Tokyo 184-8795, Japan\\
\{umeno,kitayama\}@crl.go.jp}

\isitmaketitle
\footnotetext{This work was partly supported by President's Special Research Grant of RIKEN.} 
\begin{isitabstract}
 We show that 
 chaotic spreading sequences generated by ergodic mappings of 
Chebyshev orthogonal polynomials  have better correlation 
properties for CDMA than the optimal binary sequences (Gold sequences) 
in the sense of ensemble average.  
\end{isitabstract}

\begin{isitpaper}
\isitsection{Introduction} 
Recently, the applications of chaos to practical communication systems  are gaining 
attention. Here, we investigate correlation properties of some ideal chaotic signals for their use as spreading sequences in CDMA.

\isitsection{Model and Theory}
We consider chaotic spreading sequences for CDMA as follows.
\begin{eqnarray*}
X^{(1)}_{n+1}=T[X^{(1)}_{n}]\quad\mbox{User 1}\\
\cdots \\
X^{(K)}_{n+1}=T[X^{(K)}_{n}]\quad\mbox{User K}.
\end{eqnarray*}
Here, \(T(x)\) is assumed to be one of Chebyshev polynomials
\(T_{m}(x),m\geq 2\) defined by \(T_{m}[\cos(\theta)]=\cos(m\theta)\).
\begin{equation}
  T_{1}(x)=x, T_{2}(x)=2x^{2}-1,T_{3}(x)=4x^{3}-3x,\cdots.
\end{equation}
We consider periodic sequences (\(X_{n+N}=X_{n}\)) as the spreading 
sequences. With an explicit expression of an ergodic invariant measure 
\(\rho(x)dx=\frac{dx}{\pi\sqrt{1-x^{2}}}\) of \(T_{m}(x),m\geq 2\)\cite{adler}, 
they satisfies the orthogonal relation
\begin{equation}
  \int_{-1}^{1}T_{i}(x)T_{j}(x)\rho(x)dx=\delta_{i,j}\frac{(1+\delta_{i,0})}{2}.
\end{equation}
Here, 
\(\delta_{i,j}\) is the Kronecker delta function. It is known that 
these periodic orbits of ergodic dynamical  systems are distributed according to the 
ergodic invariant measure.
Thus, we can estimate 
a mean energy of spreading sequences
\begin{equation}
\label{eq:energy}
\langle\sum_{j=1}^{N}T^{2}(X_{j})\rangle = N\int_{-1}^{1}T(x)^{2}\rho(x)dx=
\frac{1}{2}N.
\end{equation}
\begin{equation}
\label{eq:ort1}
  \langle \sum_{j=1}^{N}T(X_{j})T(X_{j+l})\rangle =N\int_{-1}^{1}T_{m}(x)T_{m^{l+1}}\rho(x)dx=0
\end{equation}
The mean interference noise \(\langle Pn \rangle\) is 0 as derived in 
\begin{equation}
\label{eq:ort2}
\langle \sum_{j=1}^{N}T(X_{j})T(Y_{j})\rangle =N\int_{-1}^{1}T(x)\rho(x)dx\cdot 
\int_{-1}^{1}T(y)\rho(y)dy=0.
\end{equation}
By the Eq. (8) of Ref. \cite{umeno99}
the mean variance of the interference noise can also be estimated as follows:
\begin{equation}
\langle Pn^{2}\rangle \equiv 
\langle [\sum_{j=1}^{N}T(X_{j})T(Y_{j})]^{2}\rangle =\frac{1}{4}N.
\end{equation}
Thus, with the use of the Gaussian assumption for \(K-1\) interference noises, we finally obtain the mean 
SNR denoted by 
\( R_{\mbox{chaos}}(K)\) as follows:
\begin{equation}
R_{\mbox{chaos}}(K)=\frac{\frac{1}{2}N}{\frac{1}{2}\sqrt{N(K-1)}}=\sqrt{\frac{N}{K-1}}.
\end{equation}
On the other hand, the mean SNR for Gold sequences obtained by
Tamura, Nakano, and  Okazaki\cite{tamura} is given by 
\begin{equation}
  R_{\mbox{Gold}}(K)=\sqrt{\frac{N^{3}}{(K-1)(N^{2}+N-1)}}.
\end{equation}
Thus, we show the following relations about mean SNR between chaotic spreading
sequences and Gold sequences. 
\begin{equation}
R_{\mbox{Gold}}(K)<R_{\mbox{chaos}}(K)\quad \mbox{for}\quad N<\infty
\end{equation}
\begin{equation}
 \lim_{N\rightarrow\infty}R_{\mbox{Gold}}(K)/R_{\mbox{chaos}}(K)=1,
\end{equation}
which exhibits a fact  that  suitable chaotic sequences 
can have better correlation properties for CDMA than the optimal binary sequences.
Our estimation of SNR for chaotic sequences is also applied to 
other chaotic maps with explicit invariant measures \cite{umeno97,umeno98,umeno_kitayama}.
\isitsection{Conclusion}
 We elucidate a correlation merit of chaotic spreading sequences by using 
the orthogonal properties of Chebyshev ergodic maps.
\isitacknowledgements
\smallskip
We thank T. Itabe and Y. Furuhama of CRL for their encouragement.
\end{isitpaper}
\begin{isitreferences}
\bibitem{adler} R. L. Adler and T. J. Rivin,{\it Proc. Am. Math. Soc.}{\bf 15}
(1964),794-796.
\bibitem{umeno99} K. Umeno,"Chaotic Monte Carlo computation: a dynamical effect of 
random-number generations", submitted to {\it Phys. Rev. }{\bf E}(1998-12)
(e-preprint: chao-dyn/9812013 at http://xxx.lanl.gov).
\bibitem{tamura} S. Tamura, S. Nakano and  K. Okazaki,{\it J. Lightwave Tech.}{\bf 3}(1985),121-127.
\bibitem{umeno97} K. Umeno,
{\it Phys. Rev. }{\bf E 55}(1997),5280-5284.
\bibitem{umeno98} K. Umeno,{\it Phys. Rev. }{\bf E 58}(1998),2644-2647.
\bibitem{umeno_kitayama}
K. Umeno and K. Kitayama,"New spreading sequences using 
periodic orbits of chaos for CDMA", submitted to 
{\it Elect. Lett.}(1998-12).
\end{isitreferences}
\end{document}